\documentclass[a4paper, table]{article}

\usepackage[utf8]{inputenc} 
\usepackage{amsmath, amsthm, amssymb, amsfonts} 
\usepackage{lipsum} 

\usepackage{graphicx} 
\usepackage{float} 
\usepackage{caption} 
\usepackage{subcaption} 

\usepackage{tabularx} 
\usepackage{tabu} 
\usepackage{booktabs} 

\usepackage[nottoc,numbib]{tocbibind} 
\usepackage[margin = 2.5cm]{geometry} 
\usepackage{microtype} 
\usepackage{titlesec} 
\usepackage{titletoc} 
\usepackage{appendix} 
\usepackage{fancyhdr} 
\usepackage[shortlabels]{enumitem} 

\usepackage{hyperref} 
\usepackage[noabbrev, capitalise]{cleveref} 

\usepackage{xcolor} 

\usepackage{tikz} 
\usetikzlibrary{positioning} 
\usetikzlibrary{calc} 

\makeatletter
\newcommand{\gettikzxy}[3]{%
  \tikz@scan@one@point\pgfutil@firstofone#1\relax
  \edef#2{\the\pgf@x}%
  \edef#3{\the\pgf@y}%
}
\makeatother

\newcommand\reporttitle{Whitepaper: Optimal Control from a \\ \bigskip Fluid Dynamics Perspective}

\newcommand\reportsubtitle{ 
}

\newcommand\groupnumber{
\textbf{}
}

\usepackage{orcidlink}
\newcommand\reportauthors{
J. Pratt \orcidlink{0000-0003-2707-3616}, M. Schneider \orcidlink{0000-0002-8505-7094}, A. Perloff \orcidlink{0000-0001-5230-0396}
\\
Decision Superiority Initiative
}

\newcommand\grouptutor{
}

\newcommand\placeanddate{
Livermore, California \today
}

\definecolor{Tue-red}{RGB}{199, 25, 24}
\definecolor{lightblue}{rgb}{.8, 1., 1.}
\definecolor{cadet}{rgb}{.3725, .619, .627}
\definecolor{cyan}{rgb}{0., .545, .545}
\definecolor{sea}{rgb}{.235, .702, .443}
\definecolor{aqua}{rgb}{.561, .737, .561}
\definecolor{turq}{rgb}{.686, .9333, .9333}
\definecolor{whiteblue}{rgb}{0.2, .8, .6}
\definecolor{bluey}{rgb}{0.2, .8, 1.}
\definecolor{ltblue4}{rgb}{.902, .957, 1}
\definecolor{seablue}{rgb}{.3725,  .619,  .627}
\definecolor{dodger}{rgb}{.0,.2758,.5151}

\hypersetup{
    colorlinks=true,
    linkcolor=dodger,
    urlcolor=dodger,
    citecolor = dodger
    }
\counterwithin{equation}{section} 
\counterwithin{figure}{section} 
\counterwithin{table}{section} 

\titleformat{\section}{\sffamily\color{dodger}\Large\bfseries}{\thesection\enskip\color{gray}\textbar\enskip}{0cm}{} 

\titleformat{\subsection}{\sffamily\color{dodger}\large\bfseries}{\thesubsection\enskip\color{gray}\textbar\enskip}{0cm}{} 

\titleformat{\subsubsection}{\sffamily\color{dodger}\bfseries}{\thesubsubsection\enskip\color{gray}\textbar\enskip}{0cm}{} 

\DeclareCaptionFont{dodger}{\color{dodger}}
\usepackage{mlmodern}

\usepackage{graphicx}
\usepackage{xspace}
\usepackage{amsmath}
\usepackage{amsfonts}
\usepackage{amssymb}
\usepackage{placeins}
\usepackage{gensymb}
\usepackage{color}
\usepackage{float}
\usepackage{hyperref}
\usepackage[english]{babel}
\usepackage{natbib}
\usepackage{dirtytalk}
\usepackage[skip=10pt plus 10pt, indent=10pt]{parskip}

\renewcommand{\vec}[1]{\mbox{\boldmath$#1$}}

\begin{document}

\begin{titlepage}

\centering

\begin{tikzpicture}

\node[opacity=0.2,inner sep=0pt,remember picture,overlay] at (4.5,-0.5){\includegraphics[width= 0.8 \textwidth]{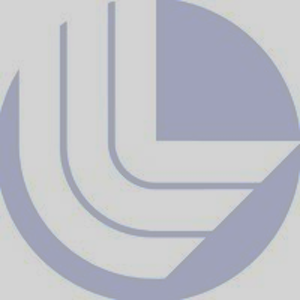}};

\node[inner sep=0pt] (logo) at (0,0)
    {\includegraphics[width=.25\textwidth]{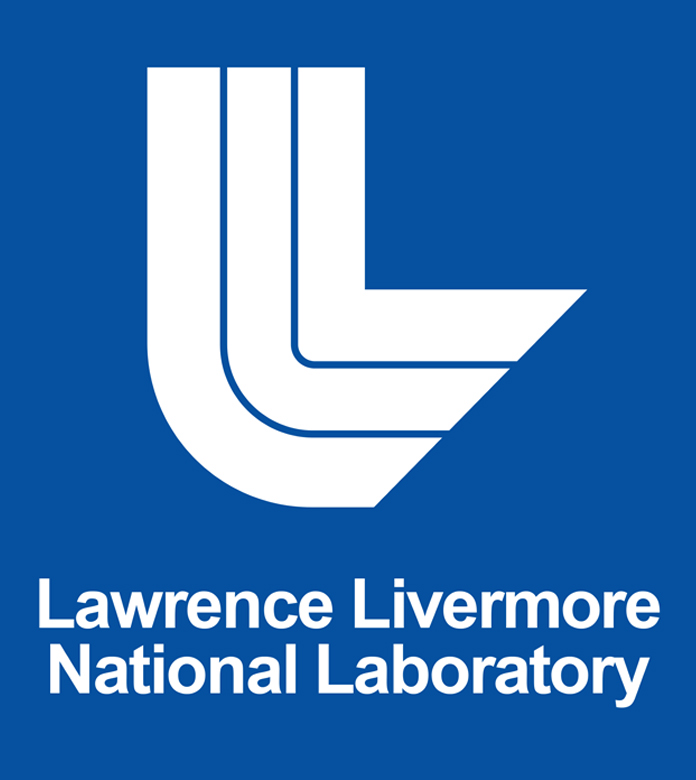}};
    
\node[text width = 0.7\textwidth, right = of logo](title){\sffamily\huge\reporttitle};

\node[text width = 0.5\textwidth, yshift = 0.75cm, below = of title](subtitle){\sffamily\Large \reportsubtitle};

\gettikzxy{(subtitle.south)}{\sffamily\subtitlex}{\subtitley}
\gettikzxy{(title.north)}{\titlex}{\titley}
\draw[line width=1mm, dodger]($(logo.east)!0.5!(title.west)$) +(0,\subtitley) -- +(0,\titley);

\end{tikzpicture}
\vspace{3cm}

\sffamily\groupnumber

\begin{table}[H]
\centering
\sffamily
\large
\begin{tabu} to 0.8\linewidth {cc}

\sffamily\reportauthors

\end{tabu}

\end{table}

\sffamily \grouptutor

\tikz[remember picture,overlay]\node[anchor=south,inner sep=0pt] at (current page.south) {\includegraphics[width=\paperwidth]{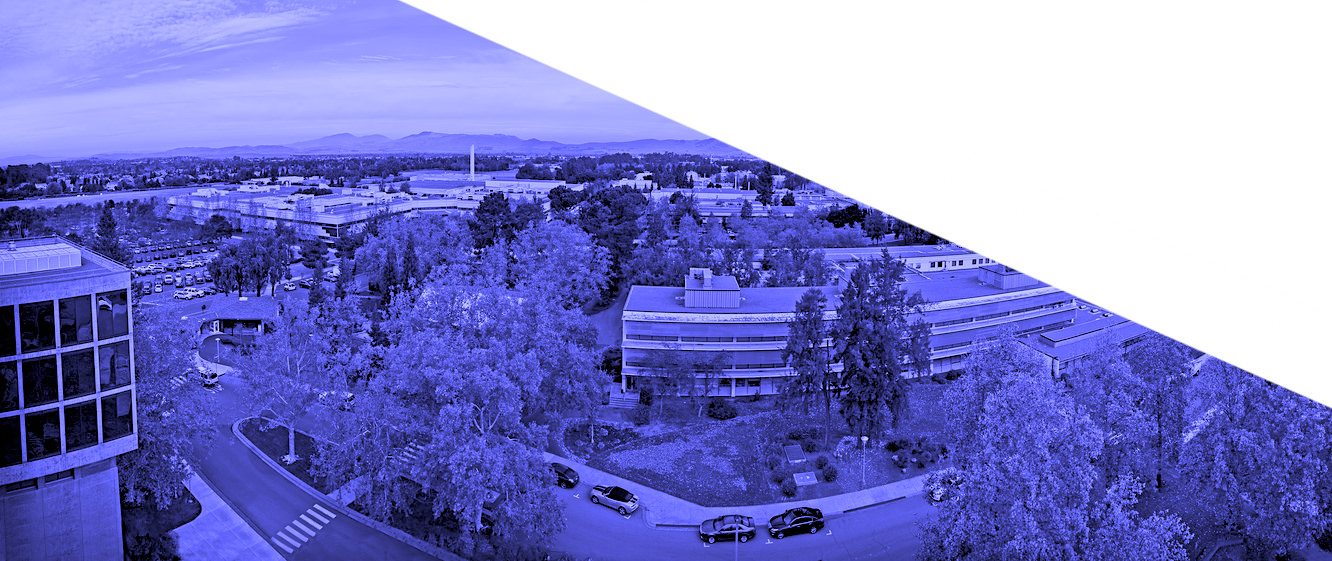}};

\mbox{}
\vfill
\sffamily \Large \textcolor{white}{\placeanddate} \\

\end{titlepage}

\newpage
\section*{Abstract}

An optimal control problem described by the Hamilton-Jacobi-Bellman equation can be developed into a problem that can be solved by general computational fluid dynamics packages.  We describe how this formulation would allow a classical problem in optimal control, Zermelo's problem, to be treated as a multi-fluid problem.  This approach has the advantage of allowing optimal navigation problems to be conducted over large areas, as well as to include moderately larger numbers of ships.  We draw comparisons between this approach and the field of fluid control for fluid animations in movies.

\vspace*{2mm}

 \pagenumbering{arabic}


\section{Introduction to the HJB equation for optimal control}

In the field of optimal control, a control is sought so that the behavior of a dynamical system optimizes an objective function.  This is most easily understood through an example.
Let us consider a problem from finance \citep{forsyth2007numerical} where an asset $S$ follows the stochastic PDE
\begin{eqnarray}\label{financestoch}
\frac{\partial S }{\partial t} = \mu S(x,t) +  \sigma S(x,t) \frac{dW}{dt}~.
\end{eqnarray}
Here $\mu$ is a drift rate, and  $\sigma$ is the volatility, a measurement of how varied (over time) the return is on an asset. The volatility is an amplitude for the time derivative of $W$, the usual Wiener process.  An optimal control problem for this system can be written for
the value $J(S,t)$ of a claim written on asset $S$ 
\begin{eqnarray}\label{hjbexample}
\frac{\partial J }{\partial t} + \sup_{q} \left( \frac{q^2 S^2}{2}J_{SS} + S J_S -rS \right)= 0~.
\end{eqnarray}
Here $r$ is the given borrowing rate, and a value of $q$ is sought such that the final selling value is specified $J(S,t_f) = D$.  This is an equation of the  Hamilton-Jacobi-Bellman (HJB)  type \citep[e.g. as discussed by ][]{kappen2005path, kuhn2019viscosity, bardi1997optimal, crespo2003stochastic}; the equation has the basic form  
\begin{eqnarray}\label{hjbform}
\frac{\partial J }{\partial t} = HJ~.
\end{eqnarray}
Here the operator $H$ is a Hamiltonian;  the Hamiltonian in an HJB equation typically has a minimum or maximum involved, making these equations notoriously difficult to solve \citep{akian2018hamilton}.  The HJB equation is typically solved backward in time, with a final time solution provided as a starting point.  HJB equations appear in probability
theory as the Kolmogorov backward equation \citep{kuhn2019viscosity}.  They thus also have a theoretical connection to the Kolmogorov forward equation, which is more commonly known as the Fokker Planck equation \citep{annunziato2014connection}.   In order to solve HJB equations,  viscosity solutions \citep{crandall2006viscosity} have been developed; these provide a mathematical framework for selecting a value function $J$ from a possibly infinite set of weak solutions. 

\citet{schneider2022optimal} examine a particular linear HJB-type equation designed to describe a linear Markov decision process (LMDP); they demonstrate that this equation can be expressed as a linear Schr\"odinger-type equation.  That transform has also been explored with mathematical rigor by \citet{teter2025control}.  In this setting, the optimal control aspect of the calculation is entirely encoded in a potential function.  We refer to this potential as the Bellman potential, and also refer to the wavefunction that moves through the Bellman potential as the Bellman wavefunction.
The transformation of a standard HJB equation to a Schr\"odinger-type equation is useful because it allows traditional scientific computing methods to be applied to optimal control.  However, dealing with a  Schr\"odinger equation also has disadvantages.  Optimal control problems are typically defined within a finite simulation volume; for this reason, the application of absorbing boundary conditions (ABC) \citep{antoine2004numerical,antoine2008review} for the Bellman wavefunction is desirable.  Because quantum mechanics allows for tunneling through barriers, the solutions obtained are likely to permit part of a Bellman wavefunction to pass through boundaries that should be impenetrable, raising questions about errors on the probability distribution.  

\section{A challenging problem for optimal control: Zermelo's navigation problem}

\begin{figure}
  \begin{minipage}[c]{0.55\textwidth}
  \resizebox{3.5in}{!}{\includegraphics{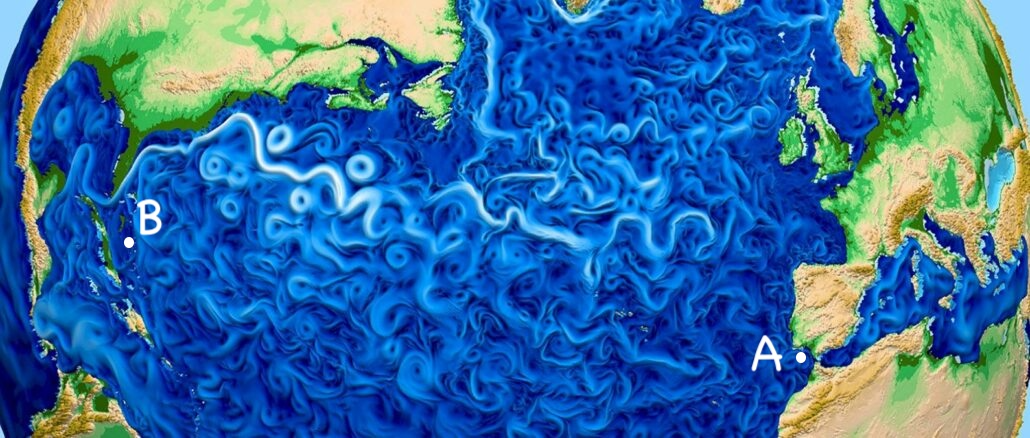}}
  \end{minipage}
  \hspace{5mm}\begin{minipage}[c]{0.4\textwidth}
        \caption{A visualization from the open-source E3SM code (image credit Argonne Leadership Computing Facility Science News) of an example setting for Zermelo's problem.  A starting point (point A) and an ending point (point B) are marked.  Optimal navigation between point A at Palos de la Frontera and point B  San Salvador would identify an optimal route for a ship moving at a given speed on a voyage across the Atlantic ocean like that of Columbus' in 1492.  \label{figzermelopath}}
  \end{minipage}
\end{figure}

Zermelo’s navigation problem  \citep{zermelo1931navigationsproblem} has become a classical problem for the field of optimal control.  The original statement of the problem is
\begin{quote}
\emph{In einer unbegrenzten Ebene, in welcher die Windverteilung durch ein
Vektorfeld als Funktion von Ort und Zeit gegeben ist, bewegt sich ein Fahrzeug mit konstanter Eigengeschwindigkeit relativ zur umgebenden Luftmasse. Wie muß das Fahrzeug
gesteuert werden, um in k\"urzester Zeit von einem Ausgangspunkte zu einem gegebenen Ziel
zu gelangen?}
\end{quote}
Typically the setting is described as a ship that is required to traverse between two points on the two-dimensional surface of a body of water in the presence of some current or wind (see Figure \ref{figzermelopath}).  In this work, we will generally refer to the current or wind as a background field.  The optimal control requirement is that the ship starting at point A should reach the prescribed final point (point B) in the least possible time; Zermelo’s problem has sometimes been reformulated to calculate an optimal energy path.  Zermelo's assumption was that the ship would be capable of a fixed speed, so that only the angle is relevant in navigating an optimal path.   A direct application of Zermelo’s problem is found in the use of active drifters as ocean sensors \citep{ouimet2014coordinated,molchanov2015active,wiggert2022navigating}.   More complex variations of Zermelo’s problem involve navigation in three-dimensional environments for active swimmers \citep{gunnarson2021learning, nasiri2022reinforcement,nasiri2024smart}, migrating animals \citep{hays2014route}, or flight \citep{techy2011optimal}.

\vspace{2mm}
When the original statement of Zermelo’s problem is modified so that background field is taken to be constant in time, it is tractable for a wide range of solution types, including machine learning techniques.  However there is no general approach that solves this problem with an evolving background field.  Thus the question is raised: how significantly does the background field need to vary for machine learning solutions to be intractable.    Zermelo's problem inherently has two different time-scales built in: one time-scale is associated with the evolution of the background field $\tau_{\mathsf{field}}$, while the other is associated with the ship $\tau_{\mathsf{ship}}$.  In the limit where the ship moves fast
\begin{eqnarray}\label{static}
\tau_{\mathsf{ship}} \ll \tau_{\mathsf{field}}~,
\end{eqnarray}
so that the background field can be assumed to be static.  In the limit where the background field evolves rapidly
\begin{eqnarray}\label{stochastic}
 \tau_{\mathsf{field}} \ll \tau_{\mathsf{ship}}~,
\end{eqnarray}
so that the background field can be assumed to be essentially stochastic, changing and then changing again before the ship can traverse further in a meaningful way.  The difficult regime is thus the regime where both ship and background field evolve on the same time-scale
\begin{eqnarray}\label{nonlinear}
 \tau_{\mathsf{field}} \sim \tau_{\mathsf{ship}}~.
\end{eqnarray}
In this case an optimal path would be determined by the ship, and begun to be traversed.  Mid-traversal, the background field would evolve so that the original path would no longer be optimal.  A  feedback loop exists between the ship and the non-linear PDEs from fluid dynamics that calculate the evolution of the background field, whether they are used to calculate a current or a wind.

A recent set of papers  \citep{biferale2019zermelo,buzzicotti2020optimal} has addressed this kind of problem using turbulence -- a setting where significant changes in the background field are implied -- and reinforcement learning methods.   They also explore a variation on  Zermelo's problem where the point B is taken to be another ship or a drifter, and so is allowed to move in time \citep{calascibetta2023optimal,calascibetta2023taming}.  \citet{biferale2019zermelo} addresses ship velocities from 20\% of the maximal flow velocity up to the maximal flow velocity.  Assuming that the background field changes on a characteristic length scale $L$, which the ship also has to traverse, we find an ordering
\begin{eqnarray}
0.2 v_{\mathsf{field}}& \leq& v_{\mathsf{ship}}  \leq v_{\mathsf{field}}~,
\\
0.2 \frac{L}{ \tau_{\mathsf{field}}} &\leq& \frac{L}{ \tau_{\mathsf{ship}} } < \frac{L}{  \tau_{\mathsf{field}} }~.
\end{eqnarray}
This translates to time scales for the ship in the range
\begin{eqnarray}
   \tau_{\mathsf{field}} <  \tau_{\mathsf{ship}} \leq 5 \tau_{\mathsf{field}}~.
\end{eqnarray}
At the point where $ \tau_{\mathsf{ship}} = 5 \tau_{\mathsf{field}} \gg \tau_{\mathsf{field}}$, the background field is approaching the stochastic situation that we mention above.
We note that it would be interesting to examine a more symmetric time range, in which the velocity of the ship could also be slightly faster than the maximal flow velocity.

 This collection of works \citep{biferale2019zermelo,buzzicotti2020optimal,calascibetta2023optimal,calascibetta2023taming} demonstrates two things: (1) that reinforcement learning based on the actor-critic algorithm can be used to solve this type of problem, i.e. the learning process produces robust and stable solutions that result in the ship moving from position A to position B, and (2) using this machine learning method is much more computationally expensive for the case of an evolving background field than for a static background field.  Methods that offer an alternative to traditional machine learning approaches, particularly methods with the potential to be more computationally efficient, are therefore desirable for the setting of background fields that evolve significantly.

\section{A Madelung Transform for the HJB equation \label{secmadelung}}

A variable transform exists that allows for the complex-valued Schr\"odinger equation to be written in the form of the Euler equations; this is called the Madelung Transform
\citep{kambe2007elementary, zylberman2022quantum, carles2012madelung,gay2020madelung,lu2015madelung}.  Here we recall the details of that transform, and interpret them in
terms of the original HJB equation for optimal control.  We begin with the basic Schr\"odinger-type equation:
\begin{eqnarray}\label{schroeq2}
i \hbar \frac{\partial }{\partial t}  \psi (r,t)  + \frac{\hbar^2}{2m}\nabla^2 \psi (r,t) - V(r) \psi (r,t) &=& 0~.
\end{eqnarray}
 To divide this equation into real and imaginary parts, without loss of generality we express the wave function in terms of a phase $\phi$ and an amplitude $A$
\begin{eqnarray}
\psi (r,t) = A(r,t) \exp{(i \phi(r,t))}~.
\end{eqnarray}
We then straightforwardly apply the chain-rule to express the time derivative of the wavefunction
\begin{eqnarray}
 \frac{\partial }{\partial t}  \psi  &=& \exp{(i \phi)} \frac{\partial }{\partial t}  A + A  \exp{(i \phi)} \frac{\partial }{\partial t} (i \phi)~,
\end{eqnarray}
Dividing by the wavefunction, we find
\begin{eqnarray}
\frac{1}{\psi} \frac{\partial }{\partial t}  \psi  &=& \frac{1}{A} \frac{\partial  A}{\partial t}  + i \frac{\partial \phi}{\partial t} ~.
\end{eqnarray}
Similarly,  we use the amplitude and phase to express the spatial gradient of the wavefunction
\begin{eqnarray}
\nabla \psi  &=&  \exp{(i \phi)}\nabla  A + A  \exp{(i \phi)} \nabla (i \phi)~,
\end{eqnarray}
and then the Laplacian
\begin{eqnarray}
\nabla^2 \psi  = \nabla \cdot \nabla \psi &=& \exp{(i \phi)}\nabla^2  A + A  \exp{(i \phi)} \nabla^2(i \phi)  
\\
&+& A  ( \exp{(i \phi)}\nabla (i \phi)) \cdot (\nabla (i \phi))  + 2 \exp{(i \phi)} (\nabla A) \cdot (\nabla (i \phi)) ~.
\end{eqnarray}
Dividing by the wavefunction, we find
\begin{eqnarray}
\frac{1}{\psi}\nabla^2 \psi  &=& \left[\frac{1}{A}\nabla^2  A 
-  (\nabla\phi)^2 \right] + i \left[\nabla^2 \phi  +  \frac{2}{A} (\nabla A) \cdot (\nabla \phi) \right]~.
\end{eqnarray}
Using these in the Schr\"odinger equation~\eqref{schroeq2} and also dividing by $\psi$, we obtain
\begin{eqnarray}
 \frac{i \hbar}{A} \frac{\partial  A}{\partial t}  - \hbar \frac{\partial \phi}{\partial t}   + \frac{\hbar^2}{2m} \left[\frac{1}{A}\nabla^2  A 
-(\nabla\phi)^2 \right] + \frac{i \hbar^2}{2m} \left[\nabla^2 \phi  +  \frac{2}{A} (\nabla A) \cdot (\nabla \phi) \right]
- V(r)   = 0~.
\end{eqnarray}
 Separating this into the real and imaginary parts we obtain two coupled equations:
\begin{eqnarray}\label{realpart}
\hbar \frac{\partial }{\partial t}  \phi - \frac{\hbar^2}{2m A}\nabla^2 A +  \frac{\hbar^2}{2m} (\nabla \phi)^2 + V(r) &=& 0~,
\\ 
 \frac{i \hbar}{A} \frac{\partial  A}{\partial t}  + \frac{i \hbar^2}{2m} \left[\nabla^2 \phi  +  \frac{2}{A} (\nabla A) \cdot (\nabla \phi) \right] &=& 0~.
\end{eqnarray}
We multiply the equation for the amplitude by $2 m A^2/i \hbar$ and apply the chain rule to the derivatives of $A$.
\begin{eqnarray}
2 mA \frac{\partial  A}{\partial t}  + \hbar A^2 \nabla^2 \phi  +  2 \hbar A (\nabla A) \cdot (\nabla \phi) &=& 0~.
\\  \label{impart}
\frac{\partial }{\partial t}  m A^2 + \hbar A^2 \nabla^2 \phi  +  \hbar \nabla A^2 \cdot \nabla \phi  &=& 0~.
\end{eqnarray}
We then identify a mass probability density
\begin{eqnarray}
\rho_{\mathsf{B}} = m \psi \psi^* = m A^2~,
\end{eqnarray}
 and a mass flux 
\begin{eqnarray}
\vec{j} = (\hbar/2i)  (\psi^* \nabla \psi - \psi \nabla \psi^*) = \hbar A^2 \nabla \phi~.
\end{eqnarray}
 A velocity field can be recovered from these two quantities 
  \begin{eqnarray}
  \vec{u}_{\mathsf{B}} = \vec{j}/\rho_{\mathsf{B}}  = \nabla \left(\frac{\hbar}{m} \phi \right) = \nabla \Phi~.  
  \end{eqnarray}
  Here in the final equality we have absorbed the constants $\hbar/m$ into a renormalized phase $\Phi$.
 The equation for the evolution of $A$ \eqref{impart} becomes
 \begin{eqnarray}
 \frac{\partial }{\partial t}  m A^2 + (m A^2) \nabla^2 (\frac{\hbar}{m}) \phi  +   \nabla m A^2 \cdot \nabla \frac{\hbar}{m} \phi  &=& 0~.
\\
 \frac{\partial }{\partial t}  \rho_{\mathsf{B}} + \nabla \rho_{\mathsf{B}} \cdot \vec{u}_{\mathsf{B}} + \rho_{\mathsf{B}} \nabla \cdot  \vec{u}_{\mathsf{B}} &=& 0~. 
\\ \label{continuityeq}
 \frac{\partial }{\partial t}  \rho_{\mathsf{B}} + \nabla \cdot ( \rho_{\mathsf{B}} \vec{u}_{\mathsf{B}} ) &=& 0~. 
\end{eqnarray}
This final equation~\eqref{continuityeq} is the usual continuity equation for a compressible fluid.

We rearrange eq.~\eqref{realpart} into an equation for the velocity.  We first divide by $m$ to obtain an equation for the renormalized phase $\Phi$:
 \begin{eqnarray}
\frac{ \hbar}{m} \frac{\partial }{\partial t}  \phi - \frac{\hbar^2}{2m^2 A}\nabla^2 A +  \frac{\hbar^2}{2m^2} (\nabla \phi)^2  &=& -\frac{ 1}{m}V(r)~.
\end{eqnarray}
We then take the gradient 
 \begin{eqnarray}
 \frac{\partial }{\partial t} \nabla \frac{ \hbar}{m} \phi - \nabla \frac{\hbar^2}{2m^2 A}\nabla^2 A + \nabla  \frac{\hbar^2}{2m^2} (\nabla \phi)^2  &=& -\nabla \frac{ 1}{m}V(r)~,
\\
 \frac{\partial }{\partial t} \nabla \Phi + (\nabla \Phi) \cdot \nabla  (\nabla \Phi)  &=& -\nabla \left[\frac{ 1}{m}V(r)+  \frac{\hbar^2}{2m^2 A}\nabla^2 A \right]~.
\end{eqnarray}
Changing the notation to that of the velocity, we find
 \begin{eqnarray}
 \label{momentumeq}
 \frac{\partial }{\partial t} \vec{u}_{\mathsf{B}} +   \vec{u}_{\mathsf{B}} \cdot \nabla \vec{u}_{\mathsf{B}} &=& - \nabla \mu ~.
\end{eqnarray}
 In the final form of this equation, we have defined a new scalar potential $\mu = V(r)/m  -  (\hbar^2/2m^2) \nabla^2 A/A $.
Eq.~\eqref{momentumeq} is the nonlinear fluid equation for the evolution of velocity.  Taken together, eqs.~\eqref{continuityeq} and \eqref{momentumeq} are an instance of the Euler equations in fluid dynamics.
 The quantity $\mu$ encapsulates all of the optimization that the Bellman potential did in the Sch\"odinger formulation -- or that the Bellman value function originally contained.  The form of this scalar potential is the same as a pressure term, a gravitational potential, or a chemical potential. We will refer to $\mu$  as the \emph{Bellman pressure}.   Using the Bellman pressure, the Bellman value function can now be interpreted as a pressure force on the flow.  

 The formulation of the HJB equation as a fluid governed by the Euler equations provides a new perspective on the field of fluid control.  The Bellman density $\rho_{\mathsf{B}}$ provides a mass-weighted probability for the Bellman wavefunction $\psi$.   Conservation of probability in the Sch\"odinger formulation of the optimal control problem thus translates to conservation of mass in the Euler fluid.  The Bellman velocity field $\vec{u}_{\mathsf{B}}$ defines the local movement of the probability density in the Sch\"odinger formulation; making this velocity explicit implies that it can be varied in either the Sch\"odinger formulation or the Euler formulation.

\section{Formulation of Zermelo's problem as multi-fluid equations}

Applying the Euler formulation in Section~\ref{secmadelung} to Zermelo's problem, we have two coupled equations that describe the optimal path of the ship
 \begin{eqnarray}
 \label{Bellmaneuler1}
 \frac{\partial }{\partial t}  \rho_{\mathsf{B}} + \nabla \cdot ( \rho_{\mathsf{B}} \vec{u}_{\mathsf{B}} ) &=& 0~, 
\\  \label{Bellmaneuler2}
 \frac{\partial }{\partial t} \vec{u}_{\mathsf{B}} +   \vec{u}_{\mathsf{B}} \cdot \nabla \vec{u}_{\mathsf{B}} &=& - \nabla \mu ~.
\end{eqnarray}
These equations are coupled to the incompressible Euler equations for the background field
 \begin{eqnarray}
 \label{Bellmaneuler3}
\nabla \cdot \vec{u}_0  &=& 0~, 
\\  \label{Bellmaneuler4}
 \frac{\partial }{\partial t} \vec{u}_0 +   \vec{u}_0 \cdot \nabla \vec{u}_0 &=& - \nabla p  ~.
\end{eqnarray}
Here we have adopted the null subscript to denote the air or fluid flow at the surface of a body of water.
Using an incompressible fluid seems like a reasonable first step to calculate the background field evolution, but compressible flows, 3D flows, and flows with additional force terms could also be examined.

For the problem as formulated in eqs.~\eqref{Bellmaneuler1} -- \eqref{Bellmaneuler4}, the use of common numerical methods for CFD could be brought to bear,  as a single online calculation.  Namely the equations for the background field eqs.~\eqref{Bellmaneuler3} -- \eqref{Bellmaneuler4} could be evolved one time step.  Then the Bellman pressure $\mu$ could be calculated based on the solution for the background field.   Finally, the equations for the ship's path eqs.~\eqref{Bellmaneuler1} -- \eqref{Bellmaneuler2} could be evolved a time step by the solver.   The ship can thus be viewed  
 as a \emph{passive} second fluid -- more complex than a passive scalar, but not significantly so.  As a Bellman fluid, the ship would initially be concentrated at a given position; it would then be transported and undergo diffusion using contributions to $\mu$ from the actual fluid.  The Bellman pressure would need to be solved for separately from the fluid equations, as is commonly done, e.g. for gravitational accelerations with a Poisson solver.  The practicality of this multi-fluid approach to optimal control depends on the efficiency of that solution, which could itself be computationally demanding.

\section{Broader implications for the field of fluid control}

 To approach the solution for the Bellman pressure, here we briefly note that this concept is parallel to ideas from the field of fluid animation.  Notable achievements in fluid animation include examples like 
 the fire from the dragon's mouth in \emph{Harry Potter and the Goblet of Fire} (Warner Bros. 2005), the ocean simulations in \emph{Moana} (Disney 2016) \citep{byun2017moana}, and  fire, explosions and water effects in \emph{Avatar: The Way of Water} (20th Century 2022) \citep{edholm2023fire,gowanlock2012particles,gowanlock2021animating}. 
  Computer graphics for fluids have to look physically plausible to the human eye, i.e. smaller length scales need to be generated by the relevant physical equations of fluid dynamics.
  However, the smallest scales, smaller than the level noticeable by the human eye, do not need to be included in fluid animation; this is similar to the philosophy of large eddy simulations (LES) \citep{schmidt2015large,lesieur1996new} where length scales smaller than some chosen cut-off are filtered out of the equations, and their effect on the large scales is modeled using a sub-grid scale model.

The existence of successful fluid animations thus demonstrates that solution of the Bellman pressure should be feasible. Most fluid animation techniques can be classified
into one of two general types:\footnote{ Deep/reinforcement learning methods have also been applied \citep{brunton2020machine}.} (1)  {proportional-derivative controllers} that guide the fluid body using additional ``ghost force'' terms that
are designed based on a distance measure between the current fluid
shape and the desired one \citep[e.g.][]{feng2018detail,thurey2009detail}, (2) {optimal controllers} that formulate the problem
as a space-time optimization over the space of possible control forces
constrained by the fluid governing equations \citep[e.g.][]{treuille2003keyframe,pan2017efficient}.   These methods operate using a concept of ``keyframes'', i.e. a sequence of states where the largest scales of the desired flow structure, e.g. the outline of the dragon's breath, is drawn into frames of the movie.  The solution from each frame to the next thus solves a problem like Zermelo's problem, where the flow has to move between a point A and a point B.
The optimization of the Bellman pressure $\mu$ required by our method echoes the fluid animation concepts of calculating external forces that move the fluid
to match a desired shape at specific times.    The details of fluid animation methods may point a way forward in the development of an algorithm for Zermelo's problem as well.

\section{Summary}

  The transformation of the HJB equation into a Schr\"odinger equation allows for the application of general scientific computing methods to be applied to otherwise intractable optimal control problems.  In this white paper, we outline a further transformation to the Euler equations which could  be convenient, especially for optimal navigation problems which could then be expressed as a multi-fluid problem.  In examining these transformations, we have drawn an analogy between the Bellman value function and a pressure-like term that we call the Bellman pressure.   We have pointed out that optimal fluid controllers used for fluid animation are  conceptually equivalent to solving for a Bellman pressure.  This suggests that such a solver would be numerically feasible.
    Using a multi-fluid set of equations, we could solve for the evolving fluid dynamics, as well as the coupled evolving optimal path of the ship using the same numerical infrastructure.
This would allow, for example, a framework for optimal path calculation to be incorporated in a code like Energy Exascale Earth System Model (E3SM) \citep{rasch2019overview} to determine optimal paths through, e.g. evolving weather patterns across the globe.

\section*{Acknowledgements}
This work was performed under the auspices of the U.S. Department of Energy by Lawrence Livermore National Laboratory under Contract DE-AC52-07NA27344.  LLNL-TR-2010440.

\bibliographystyle{unsrtnat}
\bibliography{bellman}

\end{document}